\documentclass{epl}

\title{Oblique electromagnetic instabilities for an ultra relativistic electron beam passing through a
plasma}
\author{A. Bret}

\institute{ ETSI Industriales, Universidad de Castilla-La Mancha,
13071 Ciudad Real, Spain }

\pacs{52.35.Qz}{Micro instabilities} \pacs{52.40.Mj}{Particle beam
interactions in plasmas} \pacs{52.35.Hr }{Electromagnetic waves}

\begin{document}

\maketitle

\begin{abstract}
We present an investigation of the electromagnetic instabilities
which are triggered when an ultra relativistic electron beam
passes through a plasma. The linear growth rate is computed for
every direction of propagation of the unstable modes, and
temperatures are modelled using simple waterbag distribution
functions. The ultra relativistic unstable spectrum is located
around a very narrow band centered on a critical angle which value
is given analytically. The growth rate of modes propagating in
this direction decreases like $k^{-1/3}$.
\end{abstract}

The interaction of a relativistic electron beam with a plasma is a
subject of relevance from many fields of physics ranging from
inertial confinement fusion \cite{Tabak} to some astrophysical
scenarios \cite{Dieckmann2005,Aharonian,Milosavljevic2006}. The
linear analysis of the interaction reveals an highly unstable
situation which has been investigated for a long time. Modes
propagating along the beam are unstable within a certain range of
wave-vector and form the electrostatic two-stream instability. The
so-called filamentation instability is found for modes propagating
transversely to the beam and finally, it can be proved that some
modes propagating at arbitrary angle to the beam are also unstable
\cite{fainberg}. As far as the growth rate is concerned, we see
that it is eventually a function of the parallel and perpendicular
wave vector components. As long as the beam is not relativistic,
the largest growth rate are found for wave vectors parallel to the
beam which means that the two-stream instability dominates the
linear evolution of the system in this regime \cite{BretNIMAA}.
The situation evolves  when considering  a relativistic electron
beam. Because relativistic electrons are harder to move in the
direction of their motion than in the transverse direction, the
two-stream growth rate is much more reduced than the growth rate
of the modes propagating transversely, or even obliquely. If we
denote $\gamma_b$ the beam relativistic factor, the maximum
two-stream growth rate is scaled like $\gamma_b^{-1}$, the
filamentation growth rate like $\gamma_b^{-1/2}$ whereas the
growth rate of the most unstable oblique wave vector is scaled
like $\gamma_b^{-1/3}$ \cite{fainberg}. This shows that oblique
instabilities should dominate all the more than the beam is
relativistic. The ultra relativistic limit is relevant for
astrophysical settings such as high energy cosmic rays or gamma
ray bursts production scenarios, for which some authors consider
relativistic factors up to $10^2$ \cite{Dieckmann2005} and even
$10^7$ \cite{Aharonian}. These unstable oblique modes were first
investigated in the cold limit (fluid approximation), and a
temperature dependant treatment has only been developed recently
\cite{Bret2}. As we shall see in this letter, accounting for
temperatures in the beam and the plasma results in a very narrow
oblique unstable spectrum in the ultra relativistic limit.

\begin{figure}[tbp]
\begin{center}
\includegraphics[width=0.7\textwidth]{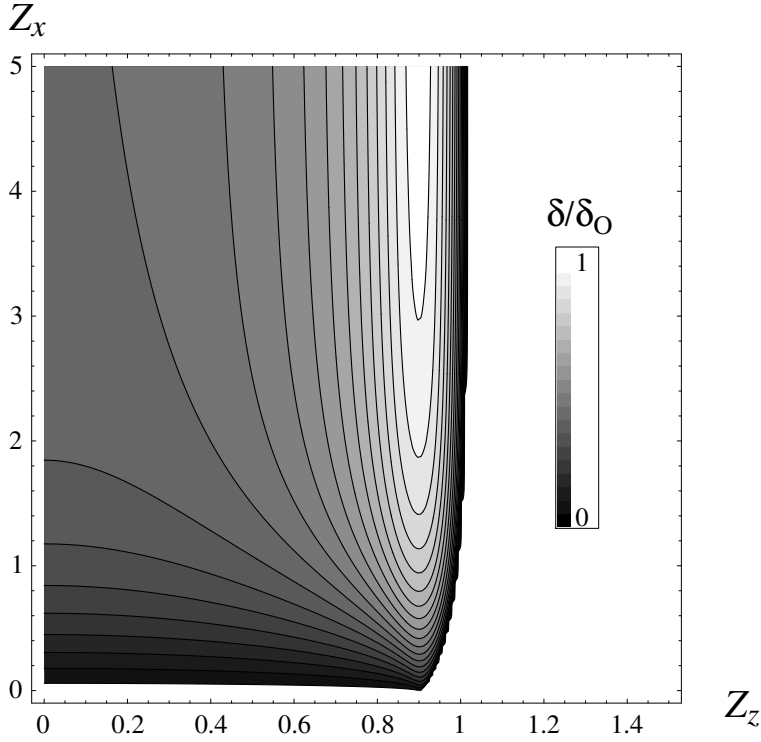}
\end{center}
\caption{Numerical evaluation of the growth rate for a cold beam
and a cold plasma, in terms of
$\mathbf{Z}=\mathbf{k}V_b/\omega_{pp}$. The growth rate is
normalized to its maximum value $\delta_O$ given by eqs.
(\ref{eq:tauxfroid}). Parameters are $n_b/n_p=0.1$ and
$\gamma_b=200$.} \label{fig:froid}
\end{figure}

We thus consider a relativistic electron beam of velocity
$\mathbf{V}_b=V_b\mathbf{e}_z$, gamma factor
$\gamma_b=1/(1-V_b^2/c^2)^{1/2}$ and density $n_b$ passing through
a plasma at electronic density $n_p\gg n_b$. Ions are supposed to
form a fixed neutralizing background and we account for a plasma
return neutralizing current \cite{Hammer} at velocity $V_p$ such
as $n_bV_b=n_pV_p$. The system is therefore charge and current
neutralized, and we study the stability of harmonic perturbations
$\propto \exp(i\mathbf{k}\cdot \mathbf{r}-i\omega)$. We implement
a 3D formalism using the linearized relativistic Vlasov equation
as well as Maxwell's ones. Given the symmetries of the problem,
the wave vector can be chosen within the $(x,z)$ plan, and the
dispersion equation reads \cite{Bret1}
\begin{equation}  \label{eq:disper}
(\omega^2\varepsilon_{xx}-k_z^2c^2)(\omega^2\varepsilon_{zz}-k_x^2c^2)-(\omega^2\varepsilon_{xz}+k_zk_xc^2)^2=0,
\end{equation}
in terms of the dielectric tensor elements
\begin{equation}  \label{eq:varepsilonIJ}
\varepsilon _{lm}= \delta _{lm}+\sum_{j=p,b}\frac{\omega
_{pj}^2}{n_j\omega^2}\int \frac{p_l}{\gamma }\frac{\partial
f_{0j}}{\partial p_m}d^{3}p
+\frac{\omega_{pj}^2}{n_j\omega ^{2}}\int \frac{p_lp_m}{\gamma^2}\frac{\mathbf{k}\cdot \partial f_{0j}/\partial \mathbf{p}%
}{m \omega -\mathbf{k}\cdot \mathbf{p}/\gamma}d^3p,
\end{equation}
where $f_{0p}$ and $f_{0b}$ are the equilibrium distribution
functions for the plasma and the beam, $m$ the electron mass,
$\omega _{pp,b}$ the electronic plasma frequencies for the plasma
and the beam, and $\gamma=(1+p^2/m^2c^2)^{1/2}$. The dispersion
equation (\ref{eq:disper}) bridges between the two-stream modes at
$k_x=0$ and the filamentation ones at $k_z=0$. Because the former
modes are longitudinal $(\mathbf{k}\parallel \mathbf{E})$ while
the later are purely transverse $(\mathbf{k}\cdot \mathbf{E}=0)$,
unstable oblique modes are neither longitudinal nor transverse and
a fully electromagnetic dispersion equation such as
(\ref{eq:disper}) is required to reach them.

If we start considering a cold mono energetic beam entering a cold
plasma, the resulting growth rate map is the one pictured on
figure \ref{fig:froid}. We plot the growth rate normalized to the
maximum oblique one in terms of the reduced wave vector
$\mathbf{Z}=\mathbf{k}V_b/\omega_{pp}$. In this cold limit, the
maximum two-stream, filamentation, and oblique growth rates read
\begin{equation}\label{eq:tauxfroid}
    \delta_{TS}\sim\frac{\sqrt{3}}{2^{4/3}}\frac{(n_b/n_p)^{1/3}}{\gamma_b},~~~
    \delta_{F}\sim\sqrt{\frac{n_b/n_p}{\gamma_b}},~~~
    \delta_{O}\sim\frac{\sqrt{3}}{2^{4/3}}\left(\frac{n_b/n_p}{\gamma_b}\right)^{1/3}.
\end{equation}
One can check that the two-stream instability is negligible
compared to the oblique and the normal modes. We also note on fig.
\ref{fig:froid} that the most unstable modes form a continuum at
$Z_z \sim 0.9$, starting from $Z_x > 3$.

The introduction of temperatures modify the picture in various
ways. As long as they remain ``small'', they can be modelled
through some simple waterbag distributions and the tensor elements
(\ref{eq:varepsilonIJ}) can be calculated exactly \cite{Bret2}. It
turns out that the resulting dispersion function has various
singularities corresponding  to various kind of preferential
coupling between a mode and some electron population. As the wave
vector is more and more oblique, these singularities evolve and
two of them overlap for a critical inclination angle $\theta_c$
with the beam, which ultra relativistic expression is simply
\begin{equation}\label{eq:angle}
    \tan\theta_c=\frac{1+n_b/n_p}{V_{tp}/V_b},
\end{equation}
where $V_{tp}$ is the plasma thermal velocity. At this junction,
let us comment what is meant by ``small temperatures''. As far as
the plasma temperature is concerned, it simply means that the
thermal velocity must remains small compared to the beam one. The
present analysis is therefore valid for non-relativistic plasma
temperatures, allowing for a 10 keV plasma for example. As far as
beam temperature is concerned, transverse and parallel
temperatures can be treated differently. Let us assume a waterbag
distribution function defining a momentum spread around
$P_b=\gamma_b m V_b$,
\begin{equation}\label{eq:distribeam}
   f_{0b}=\frac{n_b}{4P_{tb\perp}P_{tb\parallel}}\left[\Theta(pz-P_b+P_{tb\parallel})-\Theta(pz+P_b+P_{tb\parallel})\right]\left[\Theta(p_x+P_{tb\perp})-\Theta(p_x-P_{tb\perp})\right],
\end{equation}
where $\Theta$ is the step function. As can be seen on eqs.
(\ref{eq:varepsilonIJ}), the tensor elements are mostly dependent
of the $velocity$ distribution through the quantities
$\mathbf{p}/\gamma$, and this is perfectly expected since the
stability of a mode is a matter a wave particle interaction. Let
us then evaluate the velocity spread $(\Delta v_\parallel,\Delta
v_\perp)$ corresponding to the momentum spread
$(P_{tb\parallel},P_{tb\perp})$ defined above. In the present
ultra relativistic regime, we find
\begin{equation}\label{eq:velocityspread}
\Delta v_\parallel \sim \frac{1}{\gamma_b^2}\frac{\Delta
P_{tb\parallel}}{m \gamma_b},~~~\Delta v_\perp \sim
\frac{P_{tb\perp}}{m\gamma_b}.
\end{equation}
We recover for $\Delta v_\perp$ the velocity spread corresponding
to the momentum spread $P_{tb\perp}$. But $\Delta v_\parallel$ is
reduced by a factor $\gamma_b^2$  so that the parallel beam
temperature can be neglected in the very large $\gamma_b$ limit.

\begin{figure}[tbp]
\begin{center}
\includegraphics[width=0.7\textwidth]{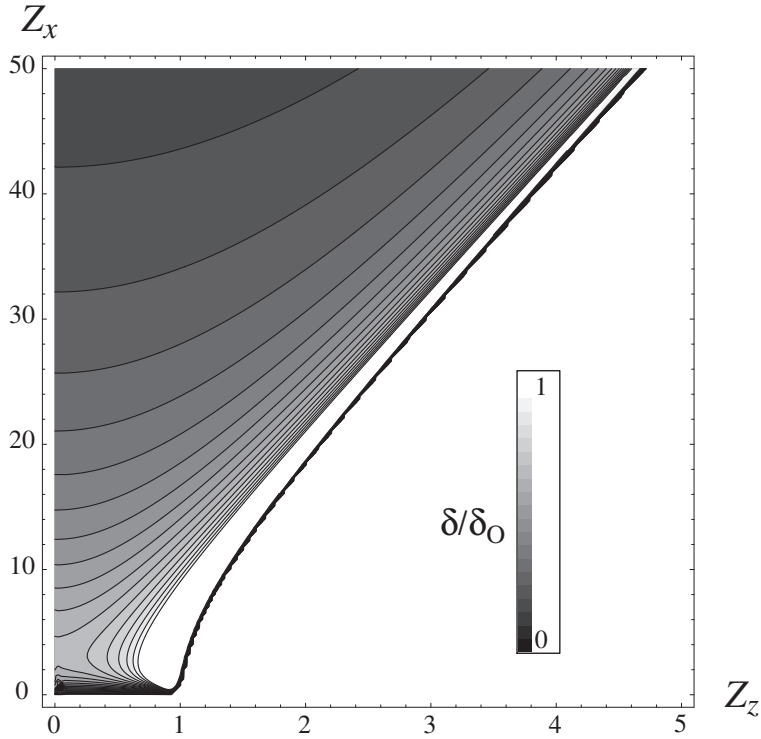}
\end{center}
\caption{Numerical evaluation of the growth rate for a hot beam
and a hot plasma, in terms of
$\mathbf{Z}=\mathbf{k}V_b/\omega_{pp}$. The growth rate is
normalized to its maximum value $\delta_O$ given by eqs.
(\ref{eq:tauxfroid}). Parameters are $n_b/n_p=0.1$ and
$\gamma_b=200$. The parallel beam velocity has been set to 0, and
every other thermal velocities are $c/10$.} \label{fig:chaud}
\end{figure}

\begin{figure}[tbp]
\begin{center}
\includegraphics[width=0.7\textwidth]{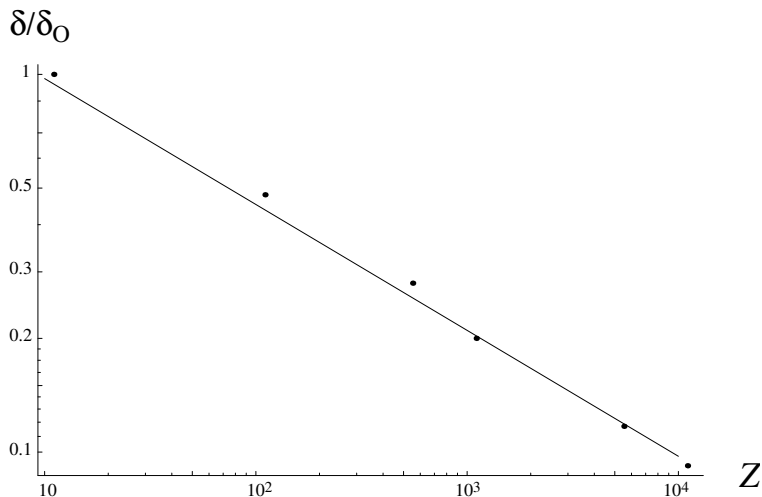}
\end{center}
\caption{Numerical evaluation of the growth rate in the critical
direction up to $Z=10^4$. Same parameters as fig. \ref{fig:chaud}.
The black points are the result of numerical calculation and the
line represents $\delta/\delta_O = 2.1Z^{-1/3}$. } \label{fig:gdZ}
\end{figure}

Considering  a relativistic beam with $\gamma_b=200$, $n_b=n_p/10$
and neglecting its parallel thermal spread, we plot on fig.
\ref{fig:chaud} the growth rate in terms of
$\mathbf{Z}=\mathbf{k}V_b/\omega_{pp}$. The continuum of the most
unstable modes observed on fig. \ref{fig:froid} has turned into a
very thin oblique region where the growth rate is higher than
$90\%$ of its cold oblique value $\delta_O$ given by eq.
(\ref{eq:tauxfroid}) up to $Z\sim 50$. Let us emphasize that
unlike fig. \ref{fig:froid} where the largest reduced wave vector
plotted is $Z_z=1.5$ and $Z_x=5$, we here extend the plot up to
$Z_z=5$ and $Z_x=50$. Modes located above the critical angle are
unstable up to $Z_x \sim \gamma_b/(V_{tp}/V_b) + Z_z \tan\theta_c$
\cite{Bret2}. We also note that the transition towards stable
modes bellow the critical angle is very sharp so that the
instability domain is almost rigourously bounded by the directions
$\theta_c$ and $\pi/2$. The critical angle obtained from eq.
(\ref{eq:angle}) is $\arctan(11)\sim 84.8^\circ$, which
corresponds very precisely to what is observed on the graph. In
order to study the growth rate at large $Z$ in the critical
direction, we plot on figure \ref{fig:gdZ} the maximum growth rate
at angle $\theta_c$ for $Z$ up to $10^4$. We observe a decrease
with $Z$ which can presently be fitted by $\delta/\delta_O = 2.1
Z^{-1/3}$. It is interesting to notice that the maximum
filamentation growth rate is here $\delta_F = 0.41 \delta_O$ so
that the growth rate in the critical direction remains larger than
the filamentation one until $Z \sim 150$.

In conclusion, we have evaluated the growth rate of the
electromagnetic instabilities for an ultra-relativistic beam
passing through a non-relativistic plasma for any angle of
propagation of the unstable modes. The parallel temperature of the
beam can be neglected, and the unstable waves are found in a
narrow domain comprised between $\pi/2$ and the critical angle
$\theta_c$ which does not depends on the beam energy (see eq.
\ref{eq:angle}). Within this region, the most unstable modes are
concentrated on a very narrow band extending around the critical
direction, with a growth rate decreasing like $Z^{-1/3}\propto
k^{-1/3}$.

\acknowledgments
 This work has been achieved under
projects FTN 2003-00721 of the Spanish Ministerio de Educaci\'{o}n
y Ciencia and PAI-05-045 of the Consejer\'{i}a de Educaci\'{o}n y
Ciencia de la Junta de Comunidades de Castilla-La Mancha. We also
thank Marie-Christine Firpo and Claude Deutsch for enriching
discussions.

\end{document}